\documentclass[sigconf, nonacm]{acmart}

\newcommand\vldbdoi{XX.XX/XXX.XX}
\newcommand\vldbpages{XXX-XXX}
\newcommand\vldbvolume{14}
\newcommand\vldbissue{1}
\newcommand\vldbyear{2020}

\newcommand\vldbtitle{\shorttitle} 
\newcommand\vldbavailabilityurl{https://github.com/audreyccheng/adrd}
\newcommand\vldbpagestyle{plain} 

\newif\ifcomments
\ifcomments
    \providecommand{\accheng}[1]{{\color{purple}{/* accheng: #1 */}}}
    \providecommand{\xs}[1]{{\color{blue}/* xshi: #1 */}}
    \providecommand{\ak}[1]{{\color{orange}/* akabcenell: #1 */}}
    \providecommand{\ion}[1]{{\color{teal}{/* ion: #1 */}}}
    \providecommand{\pb}[1]{{\color{red}{/* peter: #1 */}}}    
    \providecommand{\hng}[1]{{\color{magenta}{/* hng: #1 */}}}
    \providecommand{\lma}[1]{{\color{brown}{/* lma: #1 */}}}

\else
    \providecommand{\accheng}[1]{}
    \providecommand{\xs}[1]{}
    \providecommand{\ak}[1]{}
    \providecommand{\ion}[1]{}
    \providecommand{\pb}[1]{}
    \providecommand{\hng}[1]{}
    \providecommand{\lma}[1]{}

\fi

\newcommand{\EL}{}
\def\EL/{EL}

\usepackage{booktabs}
\usepackage{tabularx}
\usepackage{enumitem}
\usepackage[frozencache,cachedir=.]{minted}
\usepackage{subcaption}

\begin{document}
\title{AI-Driven Research for Databases}







\author{Audrey Cheng$^1$, 
Harald Ng$^2$,
Aaron Kabcenell$^3$,
Peter Bailis$^4$,
Matei Zaharia$^1$,
Lin Ma$^5$, \\
Xiao Shi$^6$,
Ion Stoica$^1$}
\affiliation{%
  \institution{$^1$UC Berkeley,
  $^2$KTH Royal Institute of Technology,
  $^3$Meta,
  $^4$Workday,
  $^5$University of Michigan,
  $^6$Maven AGI \\
  }
}

\begin{abstract}
As the complexity of modern workloads and hardware increasingly outpaces human research and engineering capacity, existing methods for database performance optimization struggle to keep pace. To address this gap, a new class of techniques, termed AI-Driven Research for Systems (ADRS), uses large language models to automate solution discovery. This approach shifts optimization from manual system design to automated code generation. The key obstacle, however, in applying ADRS is the evaluation pipeline. Since these frameworks rapidly generate hundreds of candidates without human supervision, they depend on fast and accurate feedback from evaluators to converge on effective solutions. Building such evaluators is especially difficult for complex database systems. To enable the practical application of ADRS in this domain, we propose automating the design of evaluators by co-evolving them with the solutions. We demonstrate the effectiveness of this approach through three case studies optimizing buffer management, query rewriting, and index selection. Our automated evaluators enable the discovery of novel algorithms that outperform state-of-the-art baselines (e.g., a deterministic query rewrite policy that achieves up to 6.8$\times$ lower latency), demonstrating that addressing the evaluation bottleneck unlocks the potential of ADRS to generate highly optimized, deployable code for next-generation data systems.
\end{abstract}

 \maketitle

\pagestyle{\vldbpagestyle}
\begingroup\small\noindent\raggedright\textbf{PVLDB Reference Format:}\\
Audrey Cheng, Harald Ng, Aaron Kabcenell, Peter Bailis, Matei Zaharia, Lin Ma, Xiao Shi, Ion Stoica. \vldbtitle. PVLDB, \vldbvolume(\vldbissue): \vldbpages, \vldbyear.\\
\href{https://doi.org/\vldbdoi}{doi:\vldbdoi}
\endgroup
\begingroup
\renewcommand\thefootnote{}\footnote{\noindent
This work is licensed under the Creative Commons BY-NC-ND 4.0 International License. Visit \url{https://creativecommons.org/licenses/by-nc-nd/4.0/} to view a copy of this license. For any use beyond those covered by this license, obtain permission by emailing \href{mailto:info@vldb.org}{info@vldb.org}. Copyright is held by the owner/author(s). Publication rights licensed to the VLDB Endowment. \\
\raggedright Proceedings of the VLDB Endowment, Vol. \vldbvolume, No. \vldbissue\ %
ISSN 2150-8097. \\
\href{https://doi.org/\vldbdoi}{doi:\vldbdoi} \\
}\addtocounter{footnote}{-1}\endgroup

\ifdefempty{\vldbavailabilityurl}{}{
\vspace{.3cm}
\begingroup\small\noindent\raggedright\textbf{PVLDB Artifact Availability:}\\
The source code, data, and/or other artifacts have been made available at \url{\vldbavailabilityurl}.
\endgroup
}

\section{Introduction}
Performance has long been a critical objective in database research, yet meeting the demands of modern applications is becoming increasingly difficult. The growing complexity of new workloads and hardware is quickly outpacing human research and engineering capacity. Traditionally, the design of high-performance database systems has relied on manual algorithm development. To overcome this bottleneck, recent work integrates machine learning (ML) to automate performance optimization. However, these methods largely treat the database as a black box, relying on data-intensive models either to tune existing knobs or to replace core components and perform inference at runtime. Consequently, these learned approaches remain difficult to deploy~\cite{microsoft-autotuning} due to high run-time inference overhead and unpredictable worst-case behavior.

Addressing these limitations, a new class of AI-driven techniques, termed AI-Driven Research for Systems (ADRS), is emerging to automate system design~\cite{alphaevolve,agrawal2025gepa,openevolve,cheng2025barbarians,let-the-barbarians-in}. This approach uses Large Language Models (LLMs) to generate optimized code directly rather than executing learned models at runtime, enabling a phase shift toward white-box optimization.
ADRS frameworks drive autonomous algorithm discovery, finding solutions that match or exceed the performance of state-of-the-art algorithms on various systems problems (e.g., a 13$\times$ faster load balancing algorithm~\cite{let-the-barbarians-in}). 
ADRS frameworks operate through an LLM-driven feedback loop: in each iteration, the LLM generates a candidate solution (e.g., executable code) which is evaluated against a benchmark. Promising solutions and their performance metrics are incorporated into the prompt for the subsequent iteration, guiding the search toward better programs. By evolving interpretable source code, ADRS yields solutions that are debuggable, deployable, and free of runtime inference overheads.

As the gap between modern application demands and existing database capabilities continues to grow, ADRS will be essential for system performance. Optimization was traditionally bottlenecked by the time required for researchers to manually conceptualize and implement new approaches. In contrast, ADRS can rapidly generate and evaluate many novel solutions, enabling a scale of algorithmic exploration that was previously unattainable. Thus, successfully applying ADRS to database systems is critical, as it will define how the next generation of databases is built and optimized.

Despite this transformative potential, the primary bottleneck in applying ADRS is the user-provided \textit{evaluation setup}.  In traditional research, evaluating a small number of human-designed algorithms over days or weeks is acceptable. However, ADRS can rapidly generate many solutions, which must be evaluated quickly and accurately to guide the automated optimization process. As prior work demonstrates~\cite{cheng2025barbarians,let-the-barbarians-in}, ADRS frameworks are only as effective as the feedback guiding them. Without a human in the loop to manually interpret results, the framework depends on robust evaluators that provide fast, precise, and detailed feedback to avoid ineffective optimization. Constructing this evaluation pipeline is currently a manual endeavor and a key barrier to effectively applying ADRS in practice.

Evaluation setup is especially challenging for database systems. Databases are stateful, complex, and made up of tightly coupled components, making direct evaluation prohibitively slow. The overhead of recompiling, loading data, and benchmarking each candidate solution often makes the hundreds of iterations required for convergence infeasible~\cite{cheng2025barbarians,let-the-barbarians-in,shinkaevolve}. 
Computational cost models offer a fast proxy but can fail to accurately reflect end-to-end performance~\cite{leis2015good,zhao2022queryformer,sun2019end,ding2019ai,Marcus2019Neo}. Furthermore, performance is highly workload-dependent~\cite{taobench,ramp-tao,mammoth-txns}, requiring candidate solutions to be tested against diverse benchmarks. These evaluation bottlenecks are exacerbated by massive search spaces of problems in this domain~\cite{kanellis2022llamatune} (e.g., finding the optimal set of query rewrite rules is NP-Hard~\cite{Sun2025RBot}), making discovery of effective solutions difficult within a reasonable timeframe.

To mitigate these overheads, database researchers can choose between several standard techniques to trade off between evaluation speed and quality, but applying these methods effectively requires significant manual effort.
These problem-specific methods correspond to the three key axes of evaluators: the system, workload, and search space. At the system level, researchers can build representative simulators~\cite{pbm-sampling} or performance models~\cite{index-selection} rather than executing on the real engine. To minimize workload execution time, they can evaluate on a subset of carefully selected benchmarks~\cite{taobench,oltp-bench}. Finally, they can restrict the solution search space to reduce evaluation costs at the risk of missing potential performance opportunities~\cite{smf}. Overall, these techniques navigate the evaluation trade-off but may be as difficult to apply as solving the underlying optimization problem itself (e.g., building a database simulator is notoriously difficult~\cite{lim2023database,lim2024hit}). 

\begin{figure}[t!]
  \centering
    \includegraphics[width=0.48\textwidth]{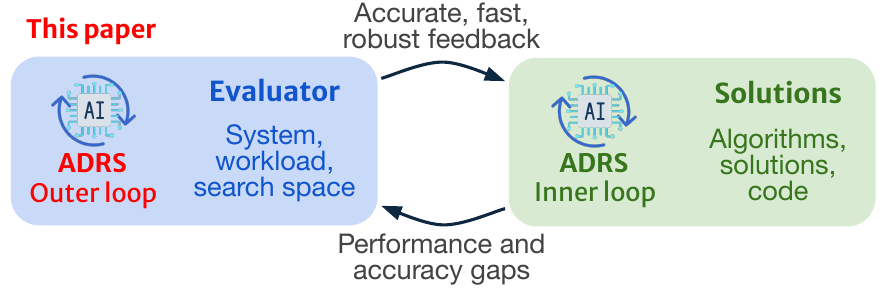}
  \vspace{-2.2em}
  \caption{We propose co-evolving the evaluator in an outer loop alongside the inner loop for solution generation.}
  \label{fig:eval-loop}
\end{figure}

In this paper, we address this critical evaluation bottleneck through \textit{co-evolution}, using LLMs to iteratively refine both candidate algorithms and their evaluation pipelines. While existing ADRS frameworks replace manual algorithm design with an automated generation loop, we extend this approach by automating the design of the evaluation components as well (Figure~\ref{fig:eval-loop}). Instead of relying on static, hand-written evaluators, our approach feeds performance signals from an inner algorithm discovery loop directly into an outer loop that continuously optimizes the evaluator. 

Since different evaluation techniques apply to different problems, we illustrate how we use AI to design efficient evaluations by considering three ``classic'' database problems that cover all evaluator axes. For the system axis, we co-evolve the simulator alongside the buffer cache policy. Guided by the insight that \textit{``more is more''} (providing multiple baselines ensures simulated gains translate to reality), we discover an algorithm achieving a 19.8\% higher hit rate and 11.4\% I/O volume savings over the state-of-the-art. For the system axis, we co-evolve an end-to-end performance model for index selection optimization. We \textit{``mind the gap''} by leveraging discrepancies between proxy metrics and E2E performance, and we discover a solution yielding up to a 6.3\% latency reduction and 2.2$\times$ lower selection time. To address the workload and search space axes, we co-evolve the workload selection and search space for query rewriting. Applying the insight to \textit{``go off what you know''} by exploiting prior empirical successes to guide evaluation, we generate a policy that reduces query latency by up to 6.8$\times$. Beyond performance, these results highlight the practical advantage of evolving white-box code: the discovered solutions are interpretable, predictable, and deployable without specialized inference infrastructure.

We argue that ADRS presents a vital step in the progression of data systems research, transitioning the field toward automatically discovered, high-performance algorithms. Our main contribution is to present an approach to automate the development of evaluators necessary for the application of ADRS. To ensure correctness, we restrict our focus to performance problems where generated solutions are easily verified without altering underlying semantics. By dynamically balancing evaluation speed and fidelity, this methodology allows us to successfully discover novel algorithms for better performance. This helps address a critical barrier in the vision for next-generation databases that can automatically adapt to changing application needs. 

In summary, we make the following contributions:
\begin{itemize}[leftmargin=*, itemsep=0pt]
\item We propose a co-evolutionary framework for ADRS to automatically develop better evaluators.
\item We detail specific methodologies to automatically navigate the trade-off between speed and quality in evaluation.
\item We demonstrate the effectiveness of this approach by applying it to three core database problems (buffer management, query rewriting, and index selection), discovering novel algorithms that exceed state-of-the-art baselines.
\end{itemize}

\section{Motivation}
In this section, we provide background on how ADRS frameworks automate the research process, shifting optimization from manual algorithm design to LLM-driven code generation. We highlight their potential for database optimization. We then discuss the critical evaluation bottleneck in applying ADRS to this domain.

\subsection{ADRS Background}
\label{subsec:background}
AI-Driven Research for Systems (ADRS)~\cite{cheng2025barbarians,let-the-barbarians-in,shinkaevolve,alphaevolve,openevolve} has emerged as a promising approach for the discovery of novel solutions. Concretely, ADRS automates two key stages of the systems research process. We define the typical workflow, which traditionally spans weeks or months, as the following five stages:
\begin{itemize}[leftmargin=*, itemsep=0pt]
    \item \textbf{Problem Formulation:} Precisely define the research scope, context, and relationship to existing work. 
    \item \textbf{Evaluation Setup:} Develop a framework to implement and measure potential solutions. Determine the appropriate evaluation objectives, workloads, and baselines.
    \item \textbf{Solution Generation:} Design a solution (e.g., an algorithm) to address the problem.
    \item \textbf{Evaluation:} Implement the solution, measure performance against workloads, and compare with baselines. If unsuccessful, the process returns to Solution Generation.
    \item \textbf{Paper Write-Up:} Document the methodology and findings.
\end{itemize}

\begin{figure}[t!]
  \centering
    \includegraphics[width=0.48\textwidth]{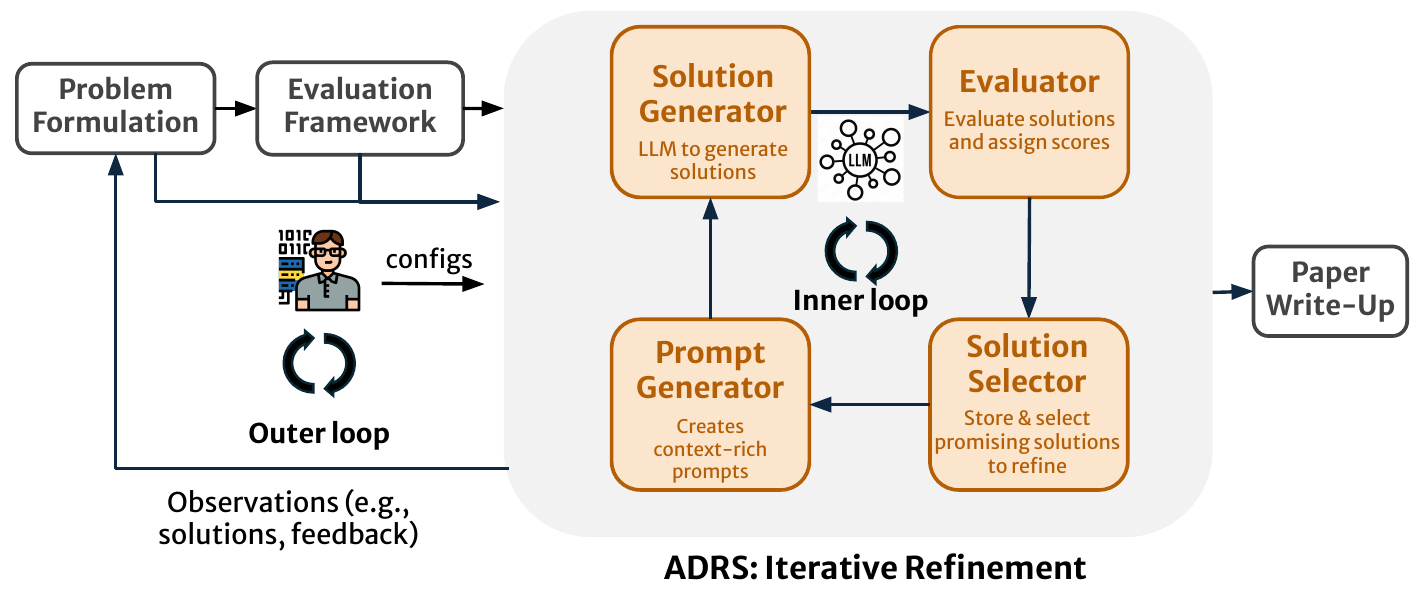}
  \vspace{-2.0em}
  \caption{ADRS implements an iterative loop to discover and refine novel solutions.}
  \label{fig:adrs-architecture}
\end{figure}

ADRS accelerates research by automating the \textbf{Solution Generation} and \textbf{Evaluation} stages. As illustrated in Figure~\ref{fig:adrs-architecture}, ADRS frameworks orchestrate an iterative loop to automate discovery, continuing until a solution meets the objectives or computational resources are exhausted. ADRS consists of five key components:
\begin{itemize}[leftmargin=*, itemsep=0pt]
\item \textbf{Prompt Generator}: Synthesizes the problem statement, system context, and prior history to guide refinement.
\item \textbf{Solution Generator}: Applies LLMs to generate or refine solutions by directly modifying code.
\item \textbf{Evaluator}: Assesses the generated solution against specified workloads to provide feedback.
\item \textbf{Storage}: Persists the history of generated solutions, execution outputs, and performance metrics.
\item \textbf{Solution Selector}: Identifies the most promising stored solutions to seed the next generation, often employing specialized evolutionary algorithms (e.g., MAPElites~\cite{alphaevolve,openevolve}).
\end{itemize}

ADRS frameworks have shown growing success on a range of problems in generating better solutions. 
Recent frameworks, including AlphaEvolve~\cite{alphaevolve}, GEPA~\cite{agrawal2025gepa}, and OpenEvolve~\cite{openevolve}, have discovered solutions that surpass state-of-the-art baselines. For instance, GEPA discovered a Mixture-of-Experts (MoE) load balancing algorithm that runs 13$\times$ faster than the best-known baseline~\cite{let-the-barbarians-in}. 
ADRS frameworks have two key strengths that enable their effectiveness. First, LLMs are trained on a vast corpus of knowledge, so they can identify patterns and techniques from diverse domains. For instance, the aforementioned MoE algorithm adapted Hamilton's Apportionment method from political science to efficiently assign replicas. Second, the automated loop enables rapid, exhaustive search of the design space. By iterating continuously until the resource budget is exhausted, ADRS can discover optimizations that human researchers might overlook.

\begin{figure}[t!]
  \centering
    \includegraphics[width=0.48\textwidth]{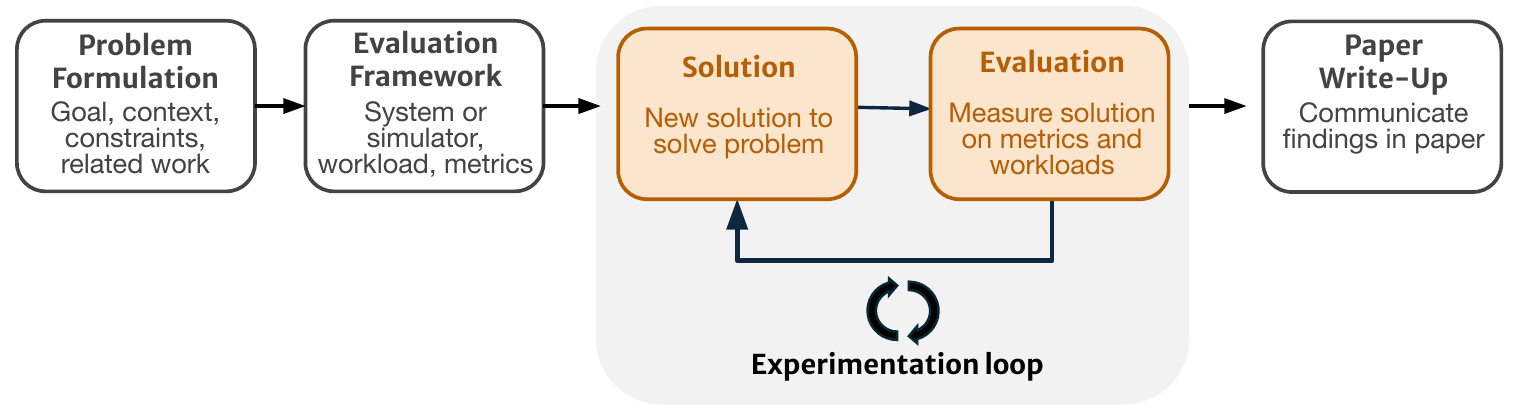}
  \vspace{-1.8em}
  \caption{The five stages of the systems research process. AI can automate the \textbf{Solution} and \textbf{Evaluation} stages (grey area).}
  \label{fig:research-process}
\end{figure}

\subsection{Potential of ADRS for Databases}
Database systems present a unique opportunity for ADRS application. Modern engines consist of highly specialized components (e.g., query optimizers, buffer managers, etc.), that rely on complex, human-designed algorithms. As workloads and hardware diversify, the complexity of manually optimizing these components increasingly outpaces human research and engineering capacity. To automate this process, the community has focused on applying learned methods to databases. These techniques treat the system as a black box, primarily centered around two approaches: (i) tuning database configurations, including knobs~\cite{kanellis2022llamatune,VanAken2017OtterTune,Zhang2022OnlineTune} and physical design (e.g., indexes~\cite{Chatterjee2022Cosine,Kraska2019SageDB,Yu2024BRAD,Ding2021MTO,ding2019ai}) with surrogate models, and more recently, LLMs~\cite{Trummer2022DBBERT,Lao2024GPTuner,Giannakouris2025LambdaTune}, and (ii) replacing specific database components (e.g., indexes~\cite{Kraska2018LearnedIndex,Ding2020ALEX,Ferragina2020PGMIndex}, query optimizers~\cite{Marcus2019Neo,Marcus2021Bao,Yang2022Balsa}, and cardinality estimators~\cite{Kipf2019MSCN,Yang2019Naru,Hilprecht2020DeepDB}) with learned models.
However, deployment of these methods has been challenging in practice due to their black-box approach. By embedding opaque models into the critical path, these approaches suffer from high run-time inference overhead, unpredictable worst-case behavior, and the need for specialized ML infrastructure.

In contrast, ADRS presents a white-box approach to optimize databases, using LLMs to generate source code that is auditable by developers, predictable in its behavior, and deployable. For instance, AlphaEvolve~\cite{alphaevolve}, an ADRS framework from DeepMind, discovered a scheduling policy that improved Google's fleet-wide utilization by 0.7\%. Beyond efficiency, this approach flexibly discovers both generalizable algorithms and specialized logic tailored to specific instances (e.g., instance-optimized systems~\cite{Kraska2019SageDB}). By automatically considering many candidate solutions, ADRS accelerates the development lifecycle and identifies optimizations that manual engineering might miss. As underlying models improve, we envision these frameworks extending beyond component optimization to synthesize entire data structures and protocols ``just-in-time'' for next-generation databases.

\subsection{The Evaluation Bottleneck}
While ADRS frameworks excel at generating solutions~\cite{cheng2025barbarians,let-the-barbarians-in}, the evaluation setup remains a critical bottleneck. These frameworks can rapidly produce many candidate solutions, which require fast and accurate evaluation to identify high-performing algorithms. Consequently, the long-term viability of ADRS requires robust evaluators, which are challenging to develop because they must carefully navigate a trade-off between evaluation speed and quality.

Developing these evaluators is especially challenging for database systems due to their complexity and large search spaces.
As components are highly coupled, testing algorithmic modifications traditionally requires running the full system end-to-end to capture intricate component interactions. While spending hours to evaluate a single human-designed algorithm was historically acceptable, LLMs now generate hundreds of candidates in minutes, shifting the bottleneck to the evaluation phase. For example, optimizing PostgreSQL's buffer replacement policy with ADRS would require modifying C source code, recompiling the engine (~3 mins), loading standard benchmarks (~30 mins for TPC-H SF=10), and executing the workload (~30 mins). With a single iteration taking over an hour, running the hundreds of cycles required for convergence would take weeks. Furthermore, evaluation can be computationally costly due to the massive search space inherent to many database performance problems. For instance, identifying optimal query rewrite rules is a combinatorial problem~\cite{Sun2025RBot}; naively applying ADRS can waste resources on ineffective combinations.

\begin{figure}[t!]
  \centering
    \includegraphics[width=0.48\textwidth]{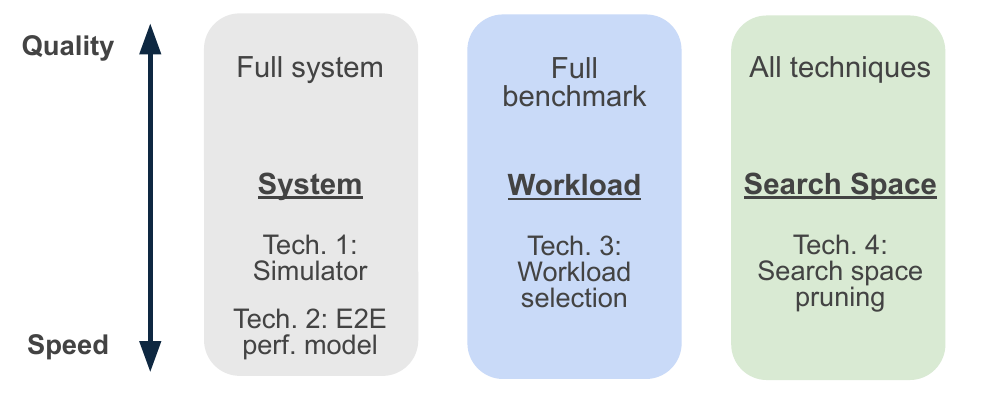}
  \vspace{-2.5em}
  \caption{Standard techniques to navigate the trade-off between evaluation speed and quality.}
  \label{fig:evaluation-tradeoff}
\end{figure}

To address these overheads, the database community has relied on established techniques that manage the fundamental trade-off between evaluation speed and quality (where quality refers to both high fidelity to end-to-end system performance and the ability to reliably converge on effective solutions). As shown in Figure~\ref{fig:evaluation-tradeoff}, these techniques target the three core components of the evaluation pipeline (system, workload, search space): 
\begin{enumerate}
    \item \textbf{Technique 1. Simulation}. Simulators avoid expensive full-system evaluation by approximating system behavior. Balancing fidelity with speed is difficult, as this requires accurately modeling complex database internals~\cite{lim2023database,lim2024hit}.
    \item \textbf{Technique 2. E2E Performance Modeling}. Modeling end-to-end behavior using simplified metrics (e.g., computational cost estimation versus query latency) enables faster results but must accurately reflect full system performance~\cite{zhao2022queryformer,sun2019end,ding2019ai,Marcus2019Neo}. 
    \item \textbf{Technique 3. Workload Selection}. Using a subset of queries accelerates benchmarking by testing fewer queries but requires careful selection to avoid overfitting and losing robust performance signals~\cite{siddiqui2022isum,xin2022locat,cai2025scompression}.
    \item \textbf{Technique 4. Search Space Pruning}. Limiting exploration to a subset of optimizations accelerates convergence but restricts coverage, requiring deep domain expertise to ensure critical opportunities are not discarded~\cite{kanellis2022llamatune}.
\end{enumerate}
These techniques can be used to achieve the fast evaluation required by ADRS, but their dependence on manual implementation creates a critical bottleneck for database systems. Traditionally, researchers have relied on intuition to navigate the speed versus quality trade-off, but applying these techniques often proves as difficult as addressing the underlying optimization problem itself.

To eliminate this bottleneck, we propose automating the evaluation setup. Just as LLMs rapidly synthesize candidate algorithms, they can dynamically generate and refine the evaluation environment. We use AI to autonomously construct simulators, develop performance models, select workloads, and prune search spaces, optimizing evaluator design in ways that were previously infeasible. In the next section, we present a co-evolutionary framework to automate key parts of the evaluation process, enabling the effective application of ADRS to databases.

\newcolumntype{L}{>{\raggedright\arraybackslash}p{\dimexpr0.4\textwidth-2\tabcolsep}}
\newcolumntype{M}{>{\raggedright\arraybackslash}p{\dimexpr0.25\textwidth-2\tabcolsep}}
\newcolumntype{R}{>{\raggedright\arraybackslash}p{\dimexpr0.35\textwidth-2\tabcolsep}}

\begin{table*}[t]
    \centering
    \caption{Summary of key insights for co-evolving the evaluator.}
    \vspace{-0.7em}
    \label{tab:case_studies}
    \begin{tabular}{LMR}
        \toprule
        \textbf{Insight} & \textbf{Case Study} & \textbf{Improvement Over SOTA} \\
        \midrule
        \textbf{Co-evolve the simulator: \textit{``More is more''}}
        \begin{itemize}[leftmargin=*, nosep]
            \item More baseline implementations and performance data enable evolution to discover algorithms that translate into real improvements.
        \end{itemize} & 
        Buffer Cache (\S\ref{sec:buffer}) 
        \begin{itemize}[leftmargin=*, nosep]
            \item Technique 1
        \end{itemize} 
        & 19.8\% higher hit rate \newline  11.4\% I/O volume savings \\ 
        \midrule
        \textbf{Co-evolve the E2E performance model: \newline \textit{``Mind the gap''}} 
        \begin{itemize}[leftmargin=*, nosep]
            \item Investigate discrepancies between proxy metrics and end-to-end performance.
        \end{itemize} & 
        Index Selection (\S\ref{sec:index}) 
        \begin{itemize}[leftmargin=*, nosep]
            \item Technique 2
        \end{itemize} & 
        6.3\% latency reduction on TPC-DS \newline 5.8\% latency reduction on TPC-H \newline 2.2$\times$ lower selection time on TPC-H \\ 
        \midrule
        \textbf{Co-evolve the workload and search space: \newline \textit{``Go off what you know''}} 
        \begin{itemize}[leftmargin=*, nosep]
            \item Exploit prior empirical successes to identify promising directions.
        \end{itemize} &
        Query Rewrite (\S\ref{sec:query-rewrite}) 
        \begin{itemize}[leftmargin=*, nosep]
            \item Technique 3
            \item Technique 4
        \end{itemize} & 
        5.4$\times$ latency reduction on TPC-H, \newline 6.8$\times$ latency reduction on DSB \\ 
        \bottomrule
    \end{tabular}
\end{table*}

\section{Co-Evolving the Evaluator}
In this paper, we demonstrate how to automate standard evaluation techniques, making ADRS feasible for databases. Our key insight is to \textit{co-evolve} the evaluation process alongside the solutions. By treating the evaluator itself as an evolvable component, the AI can dynamically navigate the speed-quality trade-offs inherent to these techniques to meet problem-specific demands.

Since different optimization problems present unique evaluation challenges, a one-size-fits-all approach is insufficient. For instance, optimizing buffer cache policies requires simulation (Technique 1) to avoid the overhead of recompilation, data loading, and benchmarking on the full system, whereas query rewrite optimization can execute directly on the full system but faces a combinatorial search space that careful workload selection (Technique 3) and search space pruning (Technique 4). As such, we eschew a monolithic architecture in favor of adapting the co-evolutionary loop to target the primary evaluation bottleneck of each problem.

Concretely, we present three case studies spanning different database components (Table~\ref{tab:case_studies}):

\textbf{Case study \# 1: buffer cache optimization (Section~\ref{sec:buffer}).} We demonstrate how to co-evolve a cache simulator alongside eviction policies to achieve a 19.8\% higher hit rate and 11.4\% I/O volume savings over the state-of-the-art baseline in PostgreSQL. This study addresses automating Technique 1 by evolving the simulator to improve fidelity so that a policy's simulated performance accurately predicts its real-world impact.  For the cache simulator, our evaluator outer loop iteratively refines the simulator's structural constraints to expose scan metadata while modeling the costs of physical I/O and lock contention. Our key insight is that ``more is more'': calibrating against multiple ground-truth baselines is essential for exposing critical system state, ensuring the simulator maintains high fidelity while preserving the expressivity required for complex policies.

\textbf{Case study \# 2: index selection optimization (Section~\ref{sec:index}).} We demonstrate how to co-evolve the performance model alongside index selection policies to achieve up to a 6.3\% reduction in query latency and 2.2$\times$ faster selection time over the SOTA baseline. This study addresses automating Technique 2 by evolving evaluation metrics towards a stable, high-fidelity fitness signal. Specifically, our evaluation outer loop iteratively designs the performance model from inaccurate computational cost estimates to an ordered execution protocol that mitigates non-deterministic noise. Our key insight is ``mind the gap'': the evaluator should investigate performance discrepancies and measurement artifacts to capture which proxy metrics matter for true end-to-end performance.

\textbf{Case study \# 3: query rewrite optimization (Section~\ref{sec:query-rewrite}).} We demonstrate how to co-evolve the evaluation workload and search space alongside query rewrite policies to achieve a 5.4$\times$ improvement in average latency on TPC-H and 6.8$\times$ on DSB over the best baseline. This study addresses automating Techniques 3 and 4 by evolving the workload subset and pruned search space to efficiently generate the missing empirical data needed for policy generation. To achieve this, our outer loop iteratively constructs highly targeted experiments to map specific optimization rules directly to matching query structures. Our key insight is ``go off what you know'': to identify promising directions without wasting cycles, the evaluator should exploit prior empirical successes.

\section{Evolving the Simulator}
\label{sec:buffer} 

In this case study, we co-evolve the simulator alongside the policy in optimizing for buffer cache performance. Our core insight is that ``more is more'': maximizing simulator fidelity requires calibration against a broader range of ground-truth baselines from the full system. By comparing against established baselines, the AI can verify what algorithmic techniques are feasible and align the simulator's performance model with actual system behavior. We demonstrate this approach by optimizing PostgreSQL's buffer manager, evolving a simulator that enables the discovery of a policy that improves hit rates by 19.8\% higher hit rate and I/O volume savings by 11.4\%.

\subsection{Performance Goal}
In this case study, we apply ADRS to evolve an optimized cache eviction algorithm. The buffer manager uses this policy to determine which database pages remain resident in the limited memory pool. We formally define the problem as follows. The input to the eviction algorithm consists of the new page request and the current cache state:
\begin{definition}
(Cache State). A cache state is a set of pages $P$ currently resident in memory, where $|P| \leq N$ and $N$ is the fixed capacity of the buffer pool. Each page $p \in P$ is associated with metadata $M_p$ (e.g., usage count, dirty flag).
\end{definition}
The output is the specific page selected for removal:
\begin{definition}
(Eviction Policy). An eviction policy is a function $f: (P, r) \rightarrow p_{victim}$ that, given the current cache state $P$ and a request for page $r \notin P$, selects a victim page $p_{victim} \in P$ to evict.
\end{definition}
Our optimization goal is to minimize the total I/O cost over a sequence of requests $R = \{r_1, r_2, \dots, r_m\}$, where a cost penalty is incurred for each cache miss ($r_i \notin P$). By maximizing the buffer cache hit rate, the evolved policy directly reduces the total volume of data read from slower secondary storage, which is the dominant cost factor in many database workloads. 
For this case study, we focus on optimizing PostgreSQL's buffer cache policy. The default eviction algorithm is Clock, which approximates LRU using a circular buffer and usage counts. To find a victim, it sweeps through pages decrementing counts until one reaches zero. While efficient, this recency-based heuristic fails on sequential scans where recently accessed pages are rarely reused. The state-of-the-art PBM-Sampling policy~\cite{pbm-sampling} approximates Belady's optimal policy by tracking active scan progress. It estimates next-access times to evict the sampled page with the furthest predicted reuse. By leveraging query semantics to increase cache hit rates, PBM-Sampling significantly reduces total I/O volume and improves hit rates on scan-heavy workloads. Ideally, we could evolve superior policies directly on top of PostgreSQL. However, running hundreds or more evolutionary iterations on PostgreSQL is impractical due to the significant overhead of full-system compilation and data loading. As such, we need a simulator to enable efficient exploration of novel algorithms.

\subsection{Evolving the Simulator via Full-System Baselines and Metrics} 
To automate policy discovery efficiently, we demonstrate how to co-evolve a simulator (Technique 1) that balances execution speed with fidelity to PostgreSQL. Our core insight is that ``more is more'': calibrating the simulator against multiple ground-truth baselines is essential to capturing critical system behaviors while still allowing enough flexibility to discover novel algorithms. Our framework iteratively refines the simulator in an evaluator outer loop based on performance of discovered policies from the solution generation inner loop. Without access to implementation logic and performance data of multiple baselines in the full system, the generator produces oversimplified simulators that constrain the search space. Access to this information guides the AI to expose rich system state (e.g., scan contexts and I/O latencies) in the simulator, enabling the discovery of policies that exploit these features for better performance.

\textbf{Naive simulation: structural constraints trap evolution.}
We initially used Cursor to generate a simulator derived from PostgreSQL's buffer cache with only the default Clock policy as context, assuming that structural fidelity would guarantee that evolved policies could be ported back to the system. Early versions mirrored internal data structures, exposing raw buffer descriptors, usage counts, and hash tables. Unfortunately, this strict replication trapped policy evolution in local optima. While PostgreSQL's buffer manager observes low-level state, it lacks global context (e.g., active scan progress) that resides in other components like the query executor. Since the simulator excluded this metadata, the solution generation inner loop could not discover strategies such as PBM-Sampling that rely on future access estimates.

\textbf{Automating simulation.}
To address this challenge, we develop a framework to co-evolve the simulator alongside the policy. Specifically, our evaluation outer loop alternates between simulator refinement and policy discovery. The loop begins by mutating simulator configurations using the full-system implementation as context to generate new variants. It then evolves policies in an inner loop using these simulated environments. Next, the framework implements and evaluates the best evolved policies in PostgreSQL, verifying correctness with unit tests. Finally, it uses this performance data to guide the refinement of the simulator.

The outer loop modifies the simulator over successive iterations to balance policy expressivity with system fidelity. Early versions relax constraints to encourage diverse solutions, but this generates low-fidelity environments that permit unrealistic techniques like full buffer pool scans. In line with our ``more is more'' insight, providing access to multiple ground-truth baselines (Clock, PBM-PQ~\cite{pbm-pq}, PBM-Sampling~\cite{pbm-sampling}) is crucial. These baselines bound the search space by demonstrating which algorithms work within PostgreSQL's architectural constraints and guide the outer loop to expose essential runtime metadata (e.g., scan context). Furthermore, by analyzing full-system performance metrics, the outer loop also discovers that modeling cache hit rate alone in the simulator is insufficient. This metric treats all evictions equally, ignoring physical I/O costs where synchronous dirty writes ($\sim$200$\mu$s) are much slower than sequential reads ($\sim$20$\mu$s).
Consequently, the outer loop accounts for these overheads and refines the simulator's scoring function (used to rank evolved policies) to include latency. This forces the inner loop to account for various performance trade-offs, driving policy evolution toward strategies that preferentially evict clean pages and sequential scan victims. Ultimately, this refinement produces our best evolved policy, which utilizes a lock-free sampling design (described in the next section).

\textbf{Final simulator.}
The final evolved simulator is trace-based and models PostgreSQL's buffer pool with realistic I/O costs (100$\mu$s for random reads, 200$\mu$s for dirty page evictions). It reports a latency score inversely proportional to simulated average I/O wait time (accounting for SSD read latency and synchronous dirty-page write cost) that represents a cost-weighted generalization of hit rate that penalizes expensive misses and dirty evictions. By incorporating essential details, such as scan tracking via block groups, ring buffer constraints, and background writer behavior, the outer loop employs iterative co-evolution to develop a suitable simulator that balances evaluation speed with fidelity.

\subsection{Evaluation}
In this section, we describe our best evolved buffer replacement policy and its results on PostgreSQL.

\subsubsection{ADRS Experimental Setup.} We implement our evaluation outer loop in Python in 5K lines of code and use OpenEvolve~\cite{openevolve} for the solution generation inner loop. We evaluate on the TPC-H workloads used the PBM-Sampling paper~\cite{pbm-sampling}, selecting a subset of the queries for evolution as a training set (our final evaluation uses the full workload). To balance fidelity and speed, our outer loop modifies the  simulator (e.g., toggling scan tracking) while the inner loop runs 50 OpenEvolve iterations to discover novel policies. Over five outer loop generations, the system evaluates and ranks simulator variants based on how well their best policies perform on PostgreSQL. We use GPT-5 and start the solution generation with the PBM-Sampling as the initial program. 

\textbf{Best evolved policy.} Our best evolved policy includes a number of key innovations compared to PBM-Sampling (Figure~\ref{fig:cache_alg_comparison}). (1) An empty-buffer fast path and lock-free sampling (lines 1--6) help bypass the scan-tracking estimator for unused buffers. This enables our policy to evaluate more candidates, improving hit rate. (2) Multi-factor scoring (lines 14--17) augments PBM-Sampling's pure Belady ranking (which considers only time-to-next-access) with bonuses for pages that have exhausted their recency protection in PostgreSQL's native Clock algorithm (i.e., pages with a \texttt{usage\_count} of 0). Relying exclusively on scan prediction fails to differentiate between recently used and stale data for point queries, so incorporating this recency penalty ensures that actively queried data is preserved even if it is not part of an ongoing sequential scan. (3) Clean/dirty tracking (lines 18--22) that maintains separate best candidates based on whether a page is unmodified (clean) or has been modified and requires writing to disk before eviction (dirty). The policy preferentially evicts clean pages, significantly reducing write-back I/O penalties that PBM-Sampling incurs by treating all evictions equally. Overall, our policy succeeds by using scan prediction to identify distant future accesses while using recency and cleanliness to break ties among eviction candidates.

\begin{figure}[t!]
\vspace{-1.4em}
  \caption{Comparison of the SOTA baseline, PBM-Sampling, with our best evolved policy.}                       
  \label{fig:cache_alg_comparison}                                                                   
  \begin{subfigure}{\linewidth}                                                                             
      \begin{minted}[
          frame=lines,
          fontsize=\small,
          linenos,
          breaklines=true
      ]{python}
  def PBM_Sampling(buffers, estimator, N=20):
      samples = []
      while len(samples) < N:
          buf = random_buffer()
          if buf.refcount != 0:
              continue  # Skip pinned buffers
          # PBM scan tracking: predict next access time
          t = estimator.estimate(buf)
          if t == NOT_REQUESTED:
              return buf  # No scan needs this page
          samples.append((buf, t))

      # Belady approximation: evict furthest access
      best = max(samples, key=lambda s: s[1])
      if best.buf.can_evict():
          return best.buf
      # Fallback: any unpinned buffer
      return random_unpinned_buffer()
        \end{minted}
      \caption{PBM-Sampling~\cite{pbm-sampling}}
      \label{fig:pbm-sampling-code}
  \end{subfigure}

  \vspace{0.5em}

  \begin{subfigure}{\linewidth}
      \begin{minted}[
          frame=lines,
          fontsize=\small,
          linenos,
          breaklines=true,
          highlightlines={2-6,14-16,18-22},
          highlightcolor=blue!10
      ]{python}
  def EvolvedPolicy(buffers, estimator, N=30):
      # Fast path: skip estimation for unused buffers
      for _ in range(3):
          buf = random_buffer()
          if buf.refcount == 0 and buf.block_group is None:
              return buf  # Empty buffer = free eviction

      best_clean = best_dirty = (None, -INF)
      while len(samples) < N:
          buf = random_unpinned_buffer()
          t = estimator.estimate(buf)
          if t == NOT_REQUESTED and not buf.is_dirty:
              return buf  # Unrequested + clean = early exit
          # Multi-factor scoring
          score = t
          if not buf.is_dirty: score += CLEAN_BONUS
          if buf.usage_count == 0: score += COLD_BONUS
          # Track best clean and dirty separately
          if buf.is_dirty:
              best_dirty = max(best_dirty, (buf, score))
          else:
              best_clean = max(best_clean, (buf, score))

      return best_clean[0] or best_dirty[0]  # Prefer clean
        \end{minted}
      \caption{Best Evolved Policy}
      \label{fig:combined-evolved-code}
  \end{subfigure}
  \vspace{-1.0em}
  \end{figure}

\subsubsection{PostgreSQL Experimental Setup.} We implement the best evolved policy in PostgreSQL 14.0. We run experiments on \texttt{c5.18xlarge} EC2 machines with 72 cores, 144 GiB of RAM, and 200GB NVMe SSD. We evaluate using two metrics: (i) buffer cache hit rate, measured via PostgreSQL's built-in statistics, and (ii) disk I/O volume. We compare performance on TPC-H~\cite{tpch}, a standard OLAP benchmark, at scale factor of 10 (\textasciitilde10GB). We follow the configurations used by PBM-Sampling paper~\cite{pbm-sampling}. We compare the performance of our best evolved policy against PBM-Sampling, PBM-Sampling full (which has additional features as detailed in the paper), and the default PostgreSQL Clock algorithm. Each data point plotted represents the average over three independent runs.

\begin{figure}[t!]
  \centering
  \includegraphics[width=0.49\textwidth]{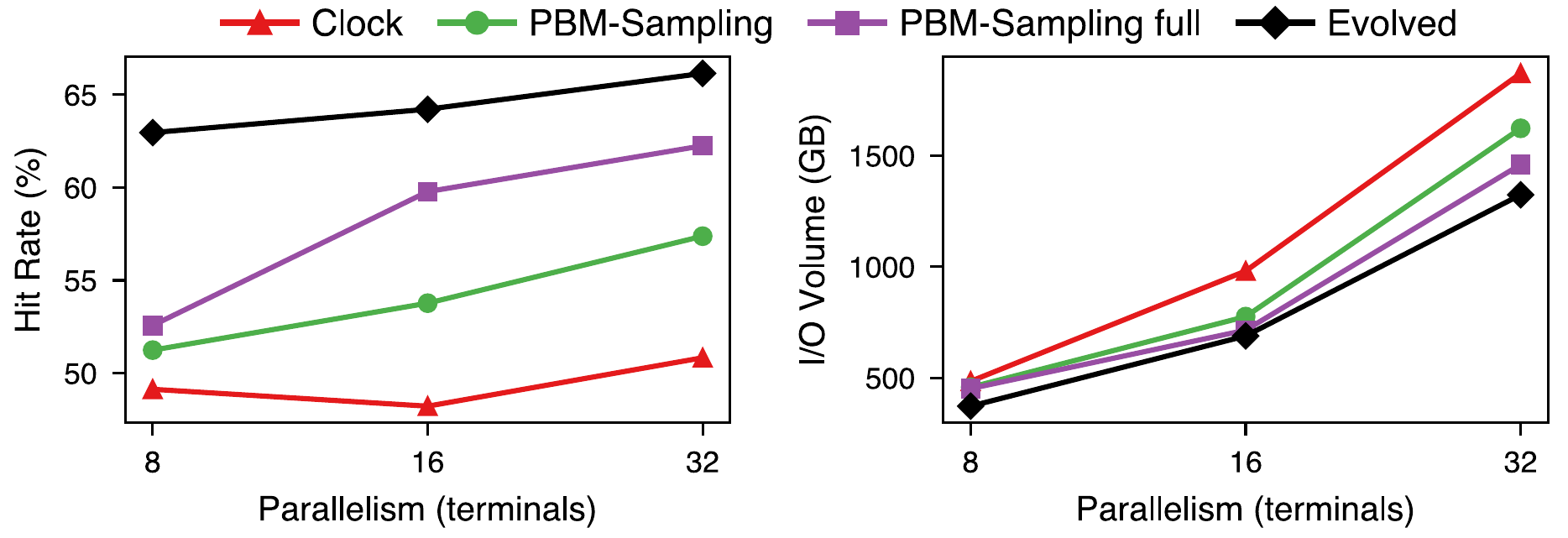}
  \vspace{-2.2em}
  \caption{Performance on TPC-H for varying parallelism.}
  \vspace{-1.5em}
  \label{fig:buffer-cache-eval}
\end{figure}

\textbf{Results.} Our evolved policy outperforms the baselines on both metrics for a range of parallel query streams (Figure~\ref{fig:buffer-cache-eval}). It achieves up to 19.8\% higher hit rate, which translates into 11.4\% I/O volume savings, compared to the best baseline. These improvements stem directly from the innovations discovered during evolution. By implementing an empty-buffer fast path, the policy reduces overhead and evaluates more eviction candidates per cycle, improving the hit rate. Furthermore, while PBM-Sampling relies solely on scan predictions and struggles with point queries, our multi-factor scoring preserves actively queried data by incorporating recency metrics. Finally, the significant I/O savings are driven by the policy's clean/dirty tracking, which preferentially evicts unmodified pages to systematically avoid the expensive write-back penalties incurred by the baseline.
\section{Evolving the Performance Model}
\label{sec:index}
In this section, we demonstrate how to co-evolve the performance model with a case study on optimizing the index selection policy. Our core insight is to ``mind the gap'': to establish a reliable fitness signal, the evaluator should actively investigate discrepancies between proxy metrics and end-to-end performance. While index selection policies are traditionally evaluated based on computational and storage cost estimates, we find this metric does not accurately capture query latency. Accordingly, we construct an evaluator outer loop that generates a reliable performance model by dynamically refining the evaluation metric to automate Technique 2. Our framework enables the discovery of an index selection policy that reduces query latency by up to 6.3\% and lowering selection time by up to 2.2$\times$.

\subsection{Optimization Goal} 
In this case study, our goal is to evolve a new index selection algorithm. Indexes accelerate data retrieval but consume storage resources and incur maintenance overhead for data modifications. Relational databases typically implement an index advisor~\cite{chaudhuri1997autoadmin}, which employs this selection policy to identify the set of indexes to use for a given workload. The input to the policy consists of a database workload, a pool of candidate indexes, and a storage limit:
\begin{definition}
(Index Selection Input). The input is a workload $W = \{q_1, q_2, \dots, q_n\}$, a set of candidate indexes $C$, and a storage budget $B$. Each index $c \in C$ has an associated storage cost $s(c)$.
\end{definition}
The output is the subset of indexes selected for creation:
\begin{definition}
(Index Selection Policy). An index selection policy $\pi: (W, C, B) \rightarrow I$ maps the input to a selected subset of indexes $I \subseteq C$ such that the total storage cost satisfies $\sum_{i \in I} s(i) \leq B$.
\end{definition}
Our optimization target is to evolve a policy $\pi$ that optimizes for two critical metrics: (1) query latency, which must be minimized to deliver user-facing performance improvements, and (2) algorithm runtime, which ensures the advisor can quickly recommend indexes in dynamic environments where workloads frequently shift, all while strictly adhering to the designated storage constraint $B$. 
Existing index selection approaches are either heuristic or learning-based, though many systems like PostgreSQL still rely on manual tuning by default since they lack a native automated advisor. A recent comprehensive study~\cite{index-selection} evaluating these approaches found significant trade-offs: while learning-based models can identify effective indexes for static queries, they suffer from high performance variance, poor robustness to data shifts, and unpredictability at larger storage budgets. Conversely, heuristic algorithms scale well and remain robust, but often fall into local optima due to their strict reliance on inaccurate optimizer computational and storage cost estimates.
Given these findings, we focus on evolving a deterministic, code-based policy that executes natively, avoiding the unpredictability of black-box models. We focus on the study's four most competitive heuristic baselines, Extend~\cite{schlosser2019extend}, DB2Advis~\cite{valentin2000db2advis}, Anytime~\cite{chaudhuri2020dta}, and AutoAdmin~\cite{chaudhuri1997autoadmin} present the optimal balance between search efficiency and estimated computation and storage cost reduction. Notably, Anytime is actively deployed in commercial production systems like Microsoft SQL Server~\cite{chaudhuri2020anytime}. 

\subsection{Evolving the Performance Model via Performance Discrepancies}
For this case study, we demonstrate how to co-evolve the performance model (Technique 2) alongside the policy itself. Our core insight is to ``mind the gap'': rather than relying on a static performance model of computational cost estimation, the evaluation outer loop investigates discrepancies between proxy metrics and end-to-end performance. 
By diagnosing the root causes of these gaps, such as measurement noise or hardware overheads ignored by the computational cost estimator, the outer loop iteratively refines the performance model used by the inner loop to generate better algorithms.

\textbf{Naive application: the cost metric illusion.} Since most prior index selection work evaluates optimizer computational cost estimates~\cite{index-selection}, we initially attempted to evolve based on this metric against standard benchmarks (TPC-H, TPC-DS). However, we found this led to ineffective policies because computational cost estimates were an inaccurate reflection of actual query latency as the model relies on static statistics and assumptions that diverge from runtime behavior. For example, on TPC-H, the DB2Advis algorithm achieves a higher computational cost reduction than Extend (49.7\% vs. 42.6\%), but degrades latency by 8\%.
This misalignment occurs because the planner’s computational cost model ignores caching (it prices every page access as if it goes disk), uses fixed I/O weights that do not match SSDs or in-memory workloads, and relies on cardinality estimates that are often wrong for correlated predicates and multi-table joins. 
Consequently, the unguided LLM search wastes evaluation cycles optimizing for theoretical improvements that failed to materialize in practice, trapping evolution in a local optimum. 

\textbf{Automating the performance model.}
To automate Technique 2, we co-evolve the performance model alongside the policy itself. The evaluation outer loop's primary goal is to construct a reliable fitness signal that ensures generated policies improve end-to-end query latency and selection time. Applying our ``mind the gap'' insight, the LLM dynamically curates the fitness function by analyzing how closely proxy metrics correlate with performance on the full system. Over multiple iterations, the AI proposes various metrics for the inner loop to optimize (e.g., a weighted hybrid of computational cost and latency, latency penalized by storage footprint). The outer loop then validates the resulting policies against their actual runtime performance. Through this process, we empirically discovered that optimizing solely for latency, once measurement noise is controlled (described below), acts as a sufficient proxy for overall system performance. This eliminates the need for complex objective functions that risk confounding the evolutionary signal.

Unlike deterministic computational cost estimates, evaluating policies based on query latency introduces significant measurement noise, a known challenge for ADRS frameworks~\cite{let-the-barbarians-in}. To establish a reliable fitness signal, the outer loop initially applied standard statistical denoising (e.g., taking the median across multiple workload executions). However, this proved insufficient: the variance between identical runs was often larger than the actual latency differences between baseline algorithms. Execution traces revealed this noise stemmed from unstable physical database states, such as volatile buffer caches, page evictions from competing queries, and interference from background processes (e.g., snapshotting). To mitigate these artifacts, the outer loop iteratively designed an ordered measurement protocol. Instead of executing the entire benchmark sequentially, which allows competing queries to thrash the cache, the framework warms up and measures each query in isolation, maintaining the exact same execution order across all evaluations.

The final performance model provides a rigorous execution pipeline that leverages latency from isolated warmup and benchmarking runs as the proxy for end-to-end performance. By exposing the full system's physical characteristics, the solution generation inner loop discovers non-obvious heuristics that contradict standard computational cost-based logic. By using workload-derived table importance, the policy identifies tables that are central to the workload's total execution time, regardless of their physical size. It applies a square-root-scaled multiplier to increase the priority of these tables, ensuring that the selection process does not overlook indexes on smaller tables that are frequently joined. This strategy improves end-to-end latency by eliminating ``micro-bottlenecks'' in complex join sequences that standard computational cost-based models often ignore in favor of large table scans.

\subsection{Evaluation}
In this section, we describe our best evolved index selection policy and its results on PostgreSQL compared to other baselines.

\subsubsection{ADRS Experimental Setup.} We implement our outer loop in 1.2K lines of Python and use OpenEvolve~\cite{openevolve} for the solution generation inner loop. To prevent overfitting, we evolve candidate policies exclusively on the TPC-DS benchmark~\cite{tpcds}. We then evaluate the final policy on both TPC-DS and TPC-H~\cite{tpch}, confirming that our performance gains successfully generalize to unseen workloads. Our evaluator enforces validity constraints, automatically disqualifying candidate policies that exceed the designated storage budget or fail to produce valid indexes. We run five iterations of the outer loop and 100 iterations of the inner loop using GPT-5 to converge on the final policy, starting the solution generation with Extend as the initial program.

\textbf{Best evolved policy.}
The best evolved index selection policy integrates several algorithmic optimizations; while the first three are known best practices for search efficiency, the policy uniquely combines them with a novel weighting strategy: (1) computational cost memoization, which caches the optimizer’s estimated workload cost for each candidate index set to ensure each is evaluated only once via hypothetical-index simulation; (2) pre-scoring by benefit/size, which ranks candidates by efficiency to drive a greedy search validated by real computational cost estimates; (3) search-space reduction, which utilizes prefix pruning, top-K seeding, and bounded extension to focus compute on the most promising candidates; and (4) table-importance weighting, a key discovery of the evolution, which uniquely boosts dimension-style tables that query planners traditionally undervalue. The policy derives this importance by summing the baseline computational costs of all queries referencing a table’s columns. It then applies a multiplier of $1.0 + \sqrt{\text{cost}_{\text{table}} / \text{cost}_{\text{max}}}$ to prioritize candidates on central tables, even when their raw benefit-to-size is modest. The square-root curve and a per-table cap ensure that the search does not disproportionately focus on a single table.

\begin{figure}[t!]
  \caption{Comparison of the Extend baseline with the best evolved index selection policy. Highlighted regions mark the key algorithmic innovations discovered by evolution.}
  \label{fig:extend_comparison}

  \begin{subfigure}{\linewidth}
      \begin{minted}[
          frame=lines,
          fontsize=\footnotesize,
          linenos,
          breaklines=true
      ]{python}
def Extend(workload, cands, eval, budget, K):
    sel, cost = [], eval.cost(workload, {})
    while len(sel) < K:
        best, best_c = None, cost
        for idx in cands:
            if idx in sel or over_budget(sel, idx, budget): continue
            c = eval.cost(workload, set(sel)|{idx})
            if c < best_c: best, best_c = idx, c
        if not best: break
        sel.append(best); cost = best_c
    for i, idx in enumerate(sel):
        for col in all_columns:
            ext = Index(idx.columns + (col,))
            if (c := eval.cost(workload, swap(sel, i, ext))) < cost: sel[i], cost = ext, c
    return set(sel)
      \end{minted}
      \caption{Extend~\cite{schlosser2019extend}}
      \label{fig:extend-baseline-code}
  \end{subfigure}

  \vspace{0.5em}

  \begin{subfigure}{\linewidth}
      \begin{minted}[
          frame=lines,
          fontsize=\footnotesize,
          linenos,
          breaklines=true,
          highlightlines={3-7,12-15,21,22,24},
          highlightcolor=blue!10
      ]{python}
def EvolvedPolicy(workload, cands, eval, budget, K):
    # (1) Cost memoization: cache each index-set evaluation
    memo = {}
    def get_cost(S):
        k = frozenset(S)
        if k not in memo:
            memo[k] = eval.cost(workload, S)
        return memo[k]
    base = get_cost({})
    q_costs = [cost_per_query(q) for q in workload]
    # (4) Table-importance weighting: sum query costs per table, boost = 1 + sqrt(t/max)
    table_cost = {}
    for q, qc in zip(workload, q_costs):
        for c in q.columns:
            table_cost[c.table] = table_cost.get(c.table, 0) + qc
    max_tc = max(table_cost.values())
    # (2) Pre-scoring by benefit/size: efficiency * table_boost
    score = {}
    for idx in cands:
        benefit = base - get_cost({idx})
        boost = 1.0 + sqrt(table_cost[idx.table()] / max_tc)
        score[idx] = (benefit / size(idx)) * boost
    # (3) Search-space reduction: prefix pruning, top-50 seeds, per-table cap
    pruned = prune_dominated_prefixes(cands, score)[:50]
    sel, cost = [], base
    for idx in pruned:
        ... # add to sel if budget, cap, min_improvement all valid
    for _ in range(4):
        for i, ex in enumerate(sel):
            for col in top_cols_by_freq(30):
                ... # extend if cost improves by min_improvement
    return set(sel)
      \end{minted}
      \caption{Best Evolved Policy}
      \label{fig:evolved-extend-code}
  \end{subfigure}
  \vspace{-1.0em}
\end{figure}

\subsubsection{Postgres Experimental Setup.} We implement the best evolved index selection policy on PostgreSQL 14~\cite{postgresql14}. We run experiments on EC2 \texttt{c5.18xlarge} EC2 machines with 72 cores, 144 GiB of RAM, and 200GB NVMe SSD. We use Benchbase~\cite{difallah2013benchbase} as our workload generator. Our experiments measure query latency, optimizer computational cost reduction, and storage utilization.
Latency is the total (frequency-weighted) execution time of the workload. For each query, we perform one warm-up run followed by a benchmark run, in interleaved order to limit cache interference. 
Each data point plotted is the median over five independent runs, with error bars representing $\pm$ 1 standard deviation.
We compare the performance of our best evolved policy against standard baselines (AutoAdmin, Anytime, DB2Advis, and Extend) on two benchmarks: TPC-H and TPC-DS. There are several queries in these benchmarks that are long-running with minimal index benefit (Q19-Q22 in TPC-H; Q0, Q5, Q56, and Q63 in TPC-DS), so we exclude them to prevent them from skewing the runtime measurements. We run all benchmarks with a 500MB index storage budget, following the index selection paper we compare against~\cite{index-selection}.

\begin{figure}[t!]
  \centering
  \includegraphics[width=0.49\textwidth]{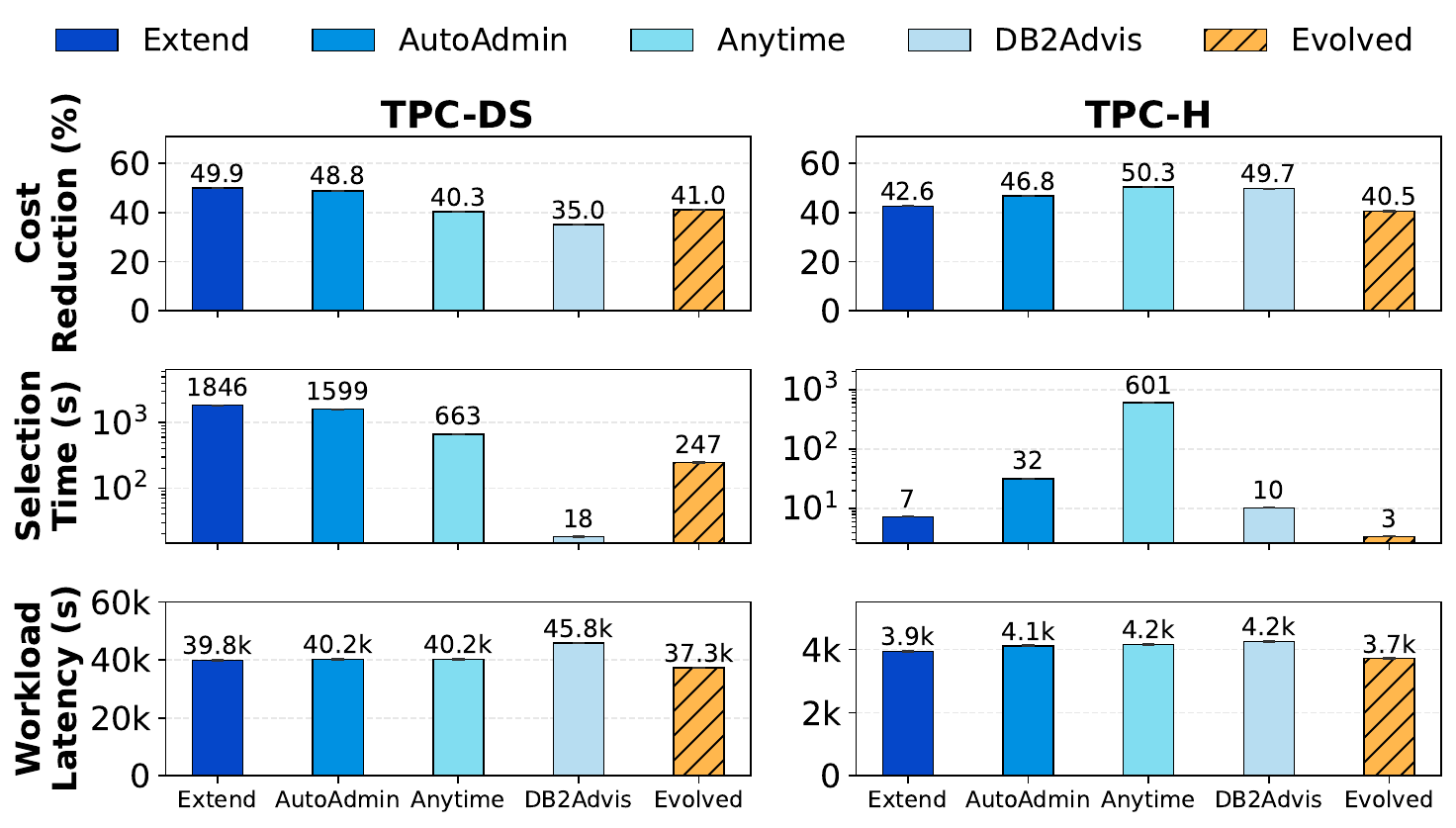}
  \vspace{-2.2em}
  \caption{Performance on TPC-DS and TPC-H.}
  \vspace{-1.5em}
  \label{fig:index-selection-eval}
\end{figure}

Our evolved algorithm reduces actual latency by 6.3\% on TPC-DS and 5.8\% on TPC-H compared to the best baseline, while achieving 2.2$\times$ lower selection time on TPC-H (3.4s vs 7.3s) and doing so within the same memory budget as the baselines (Figure~\ref{fig:index-selection-eval}).

\textbf{TPC-DS.}
Our evolved policy achieves the best latency, being 6.3\% faster than Extend (37,275 s vs 39,791 s). While it sacrifices roughly 9 percentage points of estimated computational cost reduction to achieve this, this does not impact end-to-end performance, further demonstrating that computational cost estimate is an unreliable proxy metric. This validates the performance model of denoised latency determined by our evaluation outer loop. Within the same class of iterative-search algorithms (Extend, AutoAdmin, Anytime), the evolved policy also has the fastest selection time, 7.5$\times$ faster than the baselines. DB2Advis is faster but uses a utilization-based one-pass strategy rather than an iterative search and has the worst latency among the baselines. 

\textbf{TPC-H.} As shown in Figure~\ref{fig:index-selection-eval}, our evolved policy successfully generalizes to the TPC-H benchmark, where standard computational cost-based advisors often fail to optimize for actual runtime. DB2Advis achieves a 49.7\% reduction in estimated computational cost but has the worst latency among baselines, executing 7.8\% slower than Extend. This discrepancy arises because the optimizer overestimates the benefit of indexes on high-fan-out dimension tables (e.g., date ranges) while underestimating overhead under cached execution. In contrast, our best evolved policy achieves a 5.8\% reduction in latency and the fastest selection time, improving by over 2.2$\times$ compared to the best baseline (Extend). These wins result from the inner loop discovering heuristics that directly correct the planner's computational cost-latency gap; specifically, the policy learns to reweigh candidates by their benefit-per-size, apply query computational cost weighting, and enforce table-importance boosts that prioritize true physical performance.

\section{Evolving Workload and Search} 
\label{sec:query-rewrite}
In this section, we demonstrate how to co-evolve the evaluation workload and search space with a case study on optimizing the query rewrite policy. Our core insight is ``go off what you know'': to identify promising directions, the evaluator should exploit prior empirical successes. Since unguided exploration wastes evaluation cycles, we leverage the AI to construct highly targeted experiments by automating workload selection (Technique 3) and search space pruning (Technique 4). Our framework iteratively tests these targeted combinations, analyzes rule combinations, and automatically discovers a rewrite policy that reduces average latency by up to 5.4$\times$ on TPC-H and 6.8$\times$ on DSB. 

\subsection{Performance Goal} In this case study, our goal is to evolve a query rewrite policy that reduces execution latency. In standard relational databases, the input to this policy is a SQL query, and the output is a sequence of transformation rules applied to minimize execution computational cost while preserving semantics. Formally, we define a rewrite rule to be:
\begin{definition}
(Rewrite Rule). A rewrite rule $r$ is a triplet $(c, t, f)$, where $c$ is a condition, $t$ is a transformation, and $f$ is a matching function. Given a query $q$, if $f(q)$ satisfies $c$, the transformation $t$ applies to $q$ to produce a semantically equivalent query $q' = t(q)$.
\end{definition}
We can apply a sequence of these rules on a given query:
\begin{definition}
(Rewrite Sequence). A rewrite sequence $\alpha = \langle r_1, r_2, \dots, r_k \rangle$ is an ordered list of rules selected from a rule set $R$. The sequence is valid for a query $q_0$ if each rule $r_i$ applies to the intermediate query $q_{i-1}$ to generate $q_i$. We denote the final rewritten query as $q_\alpha = q_k$.
\end{definition}
A rewrite policy maps the input query to the output sequence:
\begin{definition}
(Rewrite Policy). A rewrite policy $\pi$ is a function that maps a query $q$ to a valid rewrite sequence $\alpha$.
\end{definition}
Our objective is to minimize the execution latency of the final rewritten query, i.e., find an optimal rule sequence $\alpha^*$ such that the query execution time of $q_{\alpha^*}$ is minimized.
For this case study, we focus on optimizing Apache Calcite~\cite{apache-calcite}, a popular query rewriter, on top of PostgreSQL. PostgreSQL's native optimizer, which utilizes computational cost-based estimation and dynamic programming, serves as the default baseline. While PostgreSQL's native optimizer effectively utilizes computational cost-based estimation and dynamic programming for standard queries, it can struggle with complex analytical workloads. The state-of-the-art query rewriter, R-Bot~\cite{Sun2025RBot}, addresses this by using LLMs to guide Apache Calcite's rule-based rewriter. R-Bot retrieves relevant rewrite recipes (synthesized natural language instructions derived from Stack Overflow discussions and database forum threads) and uses an LLM to iteratively select and arrange rule combinations, achieving significant latency reductions. However, R-Bot requires runtime LLM inference and relies on an ad-hoc rule repository built from a manual selection of database documentation and other online sources. Instead, we aim to evolve a deterministic policy that executes as code during runtime.

\xs{maybe what you have discovered / what has been discovered? ``know'' implies prior expertise only before the bot loop starts}

\subsection{Evolving the Workload and Search Space via Empirical Context}
For this case study, we demonstrate how to co-evolve the evaluation workload and search space alongside the policy itself.
Our core insight is to ``go from what you know'': rather than blindly enumerating the search space, the AI uses each iteration's empirical results to decide what to test next, progressively expanding from known wins into unexplored but promising parts of the workload and search space. Existing ADRS frameworks cannot be applied directly because they lack empirical data on performant rule combinations. Accordingly, the evaluation outer loop generates this data through automated workload selection (Technique 3) and search space pruning (Technique 4).

\textbf{Naive application: exhaustive search is ineffective.} Direct application of existing ADRS frameworks to generate query rewrite policies is ineffective due to the combinatorial search space of this problem. Initially, we attempted to evolve a single global policy via LLM generation and benchmarking it against standard benchmarks (TPC-H, DSB). This approach plateaued quickly as evaluating every possible rule combination was computationally intractable. The unguided LLM search wasted evaluation cycles testing ineffective rule sequences across the entire workload, trapping evolution in a local optimum with performance far worse than the SOTA baseline.

\textbf{Automating workload selection and search space pruning.} To overcome the overheads of unguided search, our evaluation outer loop aims to efficiently generate a ground truth dataset mapping specific query features to highly performant rule sets. This empirical data directly informs the generation of an optimized rewrite policy in the solution generation inner loop. To automate Technique 3, the outer loop dynamically curates the workload by analyzing previous performance results to dictate which queries and rules to test next. The LLM curates these workloads from a designated training set of query variants. The loop applies our core insight: the AI expands from observed wins to structurally similar queries, investigates gaps where no improvement has been found, and isolates the causes of regressions. 
In each iteration, the AI outputs explicit search directives (e.g., expanding a successful rule sequence to structurally similar queries or isolating a regression source) to formulate a targeted search plan. 
For example, if the LLM notices a performance regression on queries with nested joins, it will construct latency experiments specifically for join-heavy queries using a focused subset of join-reordering rules, ensuring each evaluation cycle focuses on promising optimizations.

To automate Technique 4, the outer loop bounds the combinatorial search space through two static pruning mechanisms and a third method based on accumulated empirical evidence. First, it filters out invalid rules based on query features. For example, the system immediately discards subquery-related rewrite rules if the input query structure lacks a subquery. Second, it restricts rule orderings to pre-defined execution phases empirically discovered during an initial exploratory evaluation phase. Specifically, the outer loop requires that basic pre-processing rules (e.g., pushing down filters) be applied before complex structural transformations (e.g., converting subqueries to joins). 
Third, rule combinations that have caused regressions on similar queries are excluded from future iterations. This shrinks the search space by excluding areas likely to lead to poor performance. 
After deciding on the pruned sets of rules to apply to specific subsets of queries, the evaluation outer loop runs these experiments on PostgreSQL and records the execution latencies. Over successive iterations, this continuous testing systematically accumulates a robust ground truth dataset that maps distinct query features to performant rule sequences.

The solution generation inner loop uses the generated dataset to construct optimized rewrite policies. 
Specifically, the inner loop receives the latest dataset of execution latencies and correlates performance wins with query structure. As the first step of solution generation, it extracts discriminative features (e.g., identifying that queries with nested subqueries benefit from specific unnesting rules) to deduce which rule sequences optimize which query structures. The LLM then synthesizes these patterns into a custom, executable rewrite policy in Calcite's \texttt{HepPlanner}. If performance regressions are detected, the loop triggers a refinement cycle where the LLM investigates failures and restricts the policy logic to apply rules only when proven to improve performance. To verify correctness of generated policies, we check that every rewritten query produces the exact same rows. The final output is an optimized query rewrite policy in Calcite that uses input query features to deterministically decide which rules to apply. This design enables the database to achieve the performance benefits of extensive offline analysis with negligible runtime overhead ($<1$ms).

\subsection{Evaluation}
In this section, we describe our best evolved rewrite policy and its results on PostgreSQL compared to other baselines.

\subsubsection{ADRS Experimental Setup.} We implement our evaluation outer loop and solution generation inner loop in Python and Java, consisting of approximately 7K lines of code, and utilize Claude Sonnet 4.5. Following the R-Bot paper~\cite{Sun2025RBot}, we use the TPC-H and DSB benchmarks. To prevent overfitting, we use separate training and testing sets based on query variants. Each query template in the benchmarks has two variants with different literal values; we restrict the evaluation outer loop to only test on the first variant of each template (the training set). The inner loop yields a single, unified rewrite policy per workload (one for TPC-H and one for DSB), which is then applied to all queries (including the test set) for our final evaluation below. We run the outer loop for five iterations.


\textbf{Best evolved policy.} Our final policy includes a number of key design innovations: (1) tight feature guards that prevent regressions (e.g., explicitly disabling subquery unnesting rules if the input query contains a self-join), (2) multi-rule combinations that pair complementary rules (e.g., combining filter pushdown operations with targeted join restructuring for queries with many joins, yielding up to 10.8$\times$ speed-ups), and (3) a universal rule that automatically simplifies logic for queries with highly complex filter clauses. Furthermore, the policy orders rules effectively, consistently executing aggregate reductions before attempting complex structural transformations. 

\begin{figure}[t!]
  \centering
  \vspace{-0.7em}
  \includegraphics[width=0.48\textwidth]{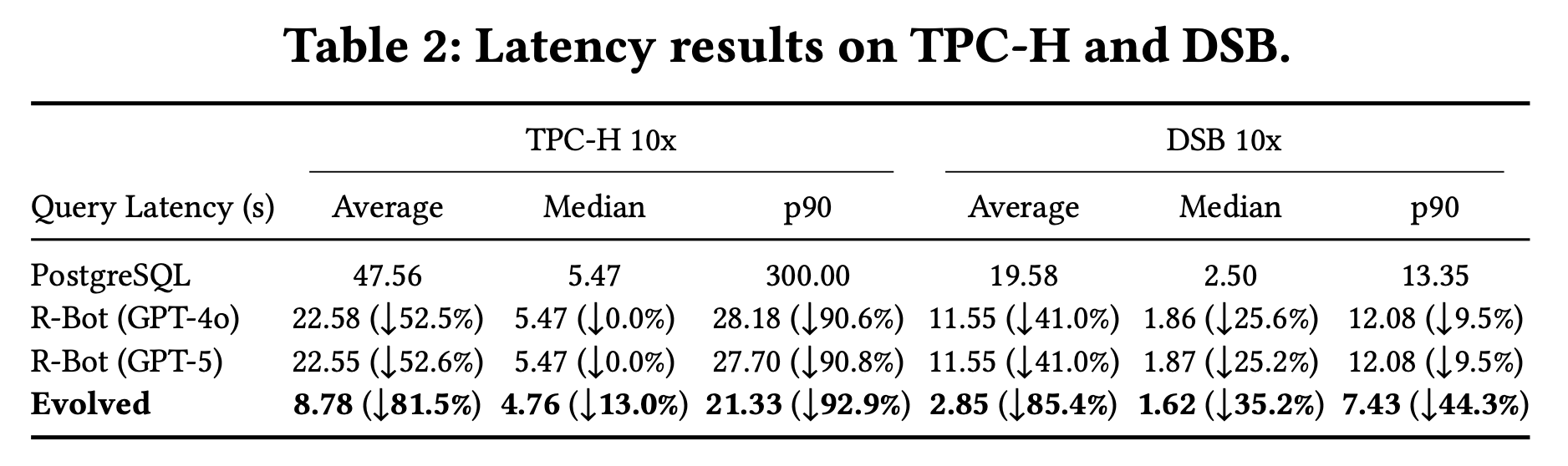}
\end{figure}


\subsubsection{PostgreSQL Experimental Setup.} We implement our best evolved rewrite policy in Apache Calcite's \texttt{HepPlanner}~\cite{apache-calcite}, integrated with PostgreSQL 17~\cite{postgresql17}. We run experiments on \texttt{c5.18xlarge} EC2 machines with 72 cores, 144 GiB of RAM, and 200GB NVMe SSD. We run each experiment three times and evaluate two metrics: (i) query latency, which measures the execution time of the rewritten query and (ii) overall latency, which captures the total time spent on both rewriting and executing the query. We compare performance on two standard analytical benchmarks: TPC-H  and DSB, both at scale factor 10. TPC-H~\cite{tpch}, a standard OLAP benchmark, consists of 22 query templates (44 instances) involving joins, subqueries, and aggregations. DSB~\cite{tpcds}, an enhanced version of TPC-DS, consists of 37 query templates (76 instances) with more complex subquery structures, unions, and window functions. We compare the performance of our best evolved policy against R-Bot (GPT-4o) and R-Bot (GPT-5), following the evaluation setup of the original paper~\cite{Sun2025RBot}.

\textbf{Query latency.} Our best evolved policy outperforms the baselines across both benchmarks (Table 2), running, on average, 2.6$\times$ faster for TPC-H and 4.0$\times$ faster for DSB compared to R-Bot. This empirical data from our evaluator outer loop enables discovery of complex rule combinations that runtime heuristics miss. For instance, our policy applies performant transformations for queries previously taking over 300s, accelerating TPC-H \texttt{query22} to 7.3s and DSB \texttt{query054} to 0.25s. In contrast, \textit{R-Bot} fails to optimize these queries since it lacks this empirical data and relies instead on PostgreSQL's runtime cost estimator. Using these proxy estimates can mislead optimization into severe performance regressions (e.g., R-Bot predicted an improvement for DSB \texttt{query100} but actually degraded execution by 19.6$\times$).

\begin{figure}[t!]
  \centering
  \vspace{-0.7em}
  \includegraphics[width=0.48\textwidth]{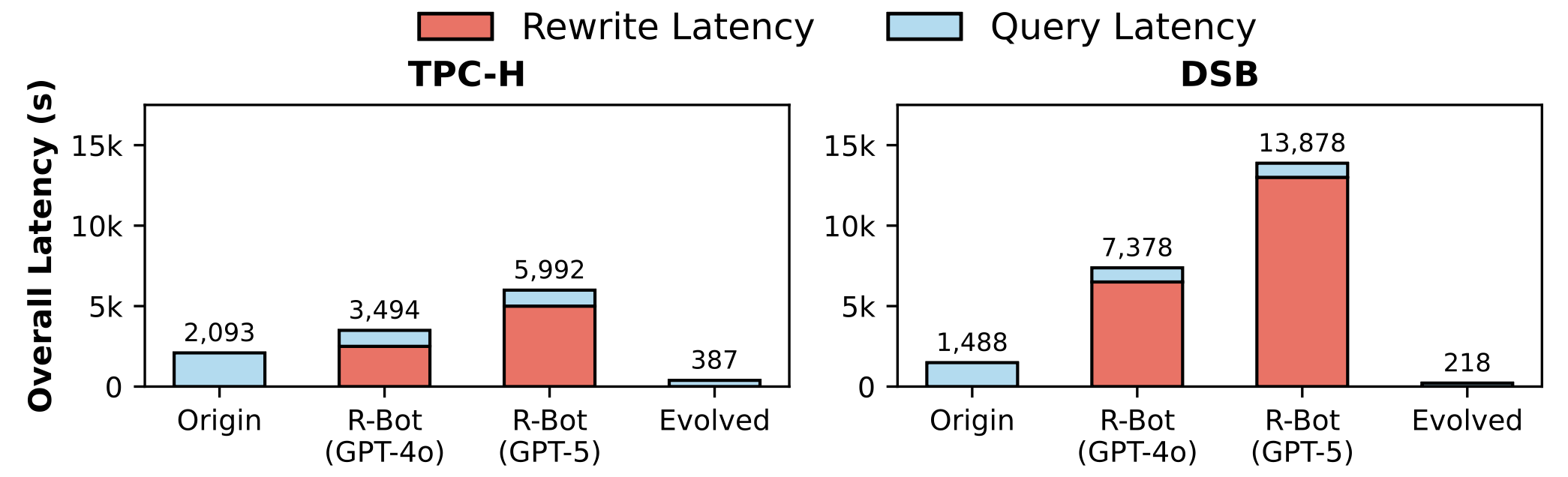}
  \caption{Overall latency on TPC-H and DSB.}
  \label{fig:qr-overall-latency}
\end{figure}

\textbf{Overall latency.} Our best evolved policy achieves the lowest overall latency (rewrite latency plus execution latency) compared to the baselines (Figure~\ref{fig:qr-overall-latency}), demonstrating the advantages of our white-box approach. Our policy executes as Java code, ensuring sub-millisecond overhead for rule selection. As such, we are, on average, 5.4$\times$ faster on TPC-H and 6.8$\times$ faster on DSB compared to PostgreSQL and 9.0$\times$ faster on TPC-H and 33.9$\times$ faster on DSB compared to R-Bot. Conversely, R-Bot invokes LLMs during runtime, incurring high inference overhead. It spends an average of over one minute per query performing RAG evidence retrieval and iterative LLM inference calls to select a rule sequence, resulting in a total workload latency that is 6.7$\times$ slower for GPT-4o and 12.9$\times$ slower for GPT-5 compared to the original baseline. For most queries, this rewrite latency completely negates any execution speedups. Furthermore, R-Bot is susceptible to LLM hallucinations, frequently analyzing a query only to hallucinate a sub-optimal or invalid rule sequence, resulting in wasted rewrite time.

\section{Discussion and Future Work} We discuss various implications of ADRS as they relate to the future of database research and design.

\textbf{Expanding beyond performance.} Current ADRS frameworks succeed primarily on performance problems since the generated algorithms preserve system semantics, simplifying verification. Applying ADRS to more complex subsystems, like concurrency control or write-ahead logging remains an open challenge. To safely evolve these critical components, evaluators must integrate rigorous correctness checks (e.g., formal verification or exhaustive fuzz testing). Future work should investigate how to automate the construction of these complex verification pipelines.

\textbf{Workload optimization.} Beyond system design, ADRS also offers significant opportunities for workload optimization, though automating this introduces a specification problem. Unlike internal system components that operate under explicit guarantees (e.g., serializability or crash recovery guarantees), developers typically express application correctness implicitly within their code. For instance, a financial app might prevent overdrafts using a mid-transaction conditional rather than a declarative database constraint. AI agents could bridge this gap by reverse-engineering formal specifications directly from the application code. With explicit rules established, ADRS frameworks could optimize queries while guaranteeing that application semantics are preserved.

\textbf{Architecting databases for AI.} The emergence of ADRS frameworks motivates a shift in database design towards predictability and modularity. Historically, developers have built highly coupled, complex architectures to maximize performance, often at the cost of system predictability~\cite{leis2015good}. However, this unpredictability hinders AI-driven search, which needs to accurately evaluate candidate solutions.
We anticipate future systems may prioritize predictable behavior over out-of-the-box efficiency to provide a foundation on which ADRS frameworks can reliably generate high-performing, workload-specific implementations.

ADRS also encourages greater modularity in systems. Future architectures may expose core components, such as the query optimizer and buffer pool, through transparent interfaces. This abstraction allows AI agents to seamlessly swap, rewrite, and evaluate individual modules, transforming the database into a platform designed for automated evolution. This structural shift aligns well with recent industry trends toward decoupled data architectures, such as the Lakebase model~\cite{lakebase}.

\textbf{The shifting role of the database researcher.} By offloading the time-consuming stages of solution generation and evaluation to an automated loop, ADRS elevates researchers to the critical task of formulating precise problems and providing these research ``assistants'' with well-defined goals and specifications. As LLMs learn to co-evolve both the solution and the evaluator, humans will increasingly act as high-level orchestrators. A key question for future research is the optimal degree of human involvement in guiding this evolutionary loop.
\vspace{-5pt}
\section{Related Work}
\textbf{Machine learning for databases.}
A large body of work applies machine learning to database systems across two main directions: (i) automated configuration and self-driving behavior, and (ii) replacing core DBMS components with learned models. For configuration, approaches frame tuning as black-box optimization guided by surrogate models, historical traces, reinforcement learning, and meta-learning~\cite{Duan2009iTuned,VanAken2017OtterTune,Zhang2019CDBTune,Li2021ResTune,Zhang2022OnlineTune,kanellis2022llamatune}. Recent work incorporates domain knowledge into the tuning loop using LLMs and agentic workflows to extract hints and generate configuration scripts~\cite{Trummer2022DBBERT,Lao2024GPTuner,Giannakouris2025LambdaTune,Li2025AgentTune,Sun2025Rabbit}. Beyond knob selection, self-driving systems broaden the scope to physical design and deployment decisions~\cite{Chatterjee2022Cosine,Kraska2019SageDB,Yu2024BRAD,Sudhir2023Pando,Ding2021MTO,pavlo2017self,ma2021mb2}.

Another line of work replaces or augments core DBMS components with learned models. This includes learned index structures that approximate data distributions~\cite{Kraska2018LearnedIndex,Ding2020ALEX,Ferragina2020PGMIndex,Wu2021LIPP,Kim2024SIA}, neural cardinality estimators~\cite{Kipf2019MSCN,Yang2019Naru,Hilprecht2020DeepDB}, and learned query optimization~\cite{Marcus2019Neo,Marcus2021Bao,Yang2022Balsa,Zhu2023Lero}. Recent work also applies ML and LLMs to query rewrite selection~\cite{Zhou2022LearnedRewrite,Li2024LLMR2,Sun2025RBot} and concurrency control policies~\cite{Wang2021Polyjuice}. In contrast to these black-box approaches, we focus on evolving optimized code with ADRS.

\textbf{AI-Driven Research.}
Recent work leverages LLMs to automate various parts of the research process. Building on genetic programming foundations~\cite{langdon2013foundations, koza1994genetic}, FunSearch~\cite{romera2024mathematical} and ELM~\cite{lehman2023evolution} use LLMs as semantic variation operators. Frameworks such as AlphaEvolve~\cite{alphaevolve} and OpenEvolve~\cite{openevolve} use the MAP-Elites algorithm and island models to evolve new algorithms, while other approaches optimize prompts and mutation operators to improve sample efficiency~\cite{agrawal2025gepa,shinkaevolve, evoprompt, pourcel2025self,wang2025thetaevolve}. Attempts at end-to-end research automation are also emerging, including open-ended code improvement (Darwin~\cite{zhang2025darwin}), agentic systems design workflows (Glia~\cite{hamadanian2025gliahumaninspiredaiautomated}), and large-scale software engineering benchmarks~\cite{nathani2025mlgym, singh2025code, fang2025comprehensive}. Furthermore, coding assistants like GitHub Copilot~\cite{GitHubCopilot}, Cursor~\cite{CursorAgent2024}, and Claude Code~\cite{ClaudeCode2025} accelerate research by enabling rapid prototyping. Recent work also explores using LLMs for performance-critical code optimization~\cite{hong2025autocompllmdrivencodeoptimization}, such as GPU kernel generation~\cite{ouyang2025kernelbench}. While most prior work focuses on automating solution generation, our focus in this paper is on co-evolving the evaluation setup.

\section{Conclusion}
This paper demonstrates that automating the evaluation pipeline unlocks the potential of AI-Driven Research for Systems (ADRS) for database systems. By co-evolving the evaluator alongside candidate solutions, we can discover novel algorithms that match or exceed state-of-the-art baselines on a range of database performance problems. While we demonstrate this approach within the context of database systems, our methodology can potentially generalize to other domains. More importantly, these results signal a fundamental shift in database research. Instead of relying on manual engineering or opaque runtime models, future optimization will increasingly depend on the generation of deployable, debuggable source code by ADRS frameworks. As application workloads and computing environments grow in complexity, these frameworks provide a scalable path for system optimization. 



\bibliographystyle{ACM-Reference-Format}
\bibliography{sample}

\end{document}